\documentclass[twocolumn,pre,superscriptaddress]{revtex4}

\usepackage{dcolumn}
\usepackage{amsmath}

\usepackage{graphicx}
\usepackage{subfigure}

\newcommand{\kin}{k^\mathrm{in}}
\newcommand{\kout}{k^\mathrm{out}}

\newcommand{\kinav}{\bar{k}^\mathrm{in}}
\newcommand{\koutav}{\bar{k}^\mathrm{out}}

\newcommand{\kappain}{\kappa^\mathrm{in}}
\newcommand{\kappaout}{\kappa^\mathrm{out}}

\renewcommand{\d}{\mathrm{d}}
\newcommand{\e}{\mathrm{e}}

\newcommand{\etal}{{\it{}et~al.}}
\newcommand{\defn}{\textit}
\newcommand{\Ord}{\mathrm{O}}

\begin{document}

\title{Random graph models for directed acyclic networks}
\author{Brian Karrer}
\affiliation{Department of Physics, University of Michigan, Ann Arbor, MI
48109}
\author{M. E. J. Newman}
\affiliation{Department of Physics, University of Michigan, Ann Arbor, MI
48109}
\affiliation{Center for the Study of Complex Systems, University of
Michigan, Ann Arbor, MI 48109}

\begin{abstract}
  We study random graph models for directed acyclic graphs, an important
  class of networks that includes citation networks, food webs, and
  feed-forward neural networks among others.  We propose two specific
  models, roughly analogous to the fixed edge number and fixed edge
  probability variants of traditional undirected random graphs.  We
  calculate a number of properties of these models, including particularly
  the probability of connection between a given pair of vertices, and
  compare the results with real-world acyclic network data finding that
  theory and measurements agree surprisingly well---far better than the
  often poor agreement of other random graph models with their
  corresponding real-world networks.
\end{abstract}

\maketitle

\section{Introduction}
\label{sec:intro}

A \defn{directed acyclic graph} is a directed graph with no cycles---closed
paths across the graph that start and end at the same vertex and follow
edges only in their forward direction.  Directed acyclic graphs are a
fundamental class of networks that occur widely in natural and man-made
settings.  The best-studied examples are citation networks, networks in
which the vertices represent documents and the directed edges represent
citations between them.  Citation networks of learned papers have long been
an object of study in the information sciences~\cite{Price65,ER90,Seglen92}
and more recently in physics~\cite{Redner98,LLJ03}, and citation networks
of patents~\cite{JT02} and legal cases~\cite{Fowler07,LCSN07} have also
received some attention in the last few years.  Directed acyclic graphs
occur in many other areas too.  In biology, phylogenetic networks
representing gene transfer are strictly acyclic and food webs are
approximately so.  In computer science and engineering acyclic or
approximately acyclic graphs occur in data structures, software call
graphs, and feed-forward neural networks.  In pure mathematics acyclic
graphs are studied for their own sake~\cite{PT01,BE84,McKay04} and as a
representation of partially ordered sets~\cite{Luczak91} and random graph
orders~\cite{AF89,BB97}, while in statistics the widely used Bayesian
networks are an acyclic graph version of probabilistic graphical
models~\cite{Jensen01,IC02,MWR06}.

Over the years, the study of networks has been substantially illuminated by
the development of random graph models.  Such models include the original
(Poisson) random graph famously studied by Erd\H{o}s and
R\'enyi~\cite{ER59,ER60}, the configuration model of Molloy and Reed and
others~\cite{Bollobas80,Luczak92,MR95,MR98} and its generalizations to
directed, bipartite, and other network types~\cite{NSW01,DMS01a}, the
small-world model of Watts and Strogatz~\cite{WS98}, exponential random
graphs~\cite{HL81,Strauss86}, and others.  These models, combining simple
definitions with complex but still analytically accessible structures, have
provided an invaluable window on the expected behavior of large networks,
as well as serving as the starting point for many other more sophisticated
models and calculations.

To the best of our knowledge, however, no corresponding model has been
studied for directed acyclic graphs---no equivalent of the configuration
model for networks such as citation networks or food webs.  In this paper,
we propose such a model and study its properties in detail, giving
derivations of a variety of quantities of interest, extensive numerical
simulations, and comparisons with the behavior of real-world acyclic
graphs, with which, in some cases, the model appears to be in surprisingly
good agreement.  A brief report of some of the material in this paper has
appeared previously as Ref.~\cite{Karrer09a}.

\section{Acyclic graphs and ordered graphs}
\label{sec:ordered}

To correctly specify a random graph model for directed acyclic graphs it is
crucial first to understand the reason why such graphs are acyclic in real
life.  In most practical examples the acyclic nature of the network arises
because the vertices are ordered.  In citation networks and phylogenetic
networks, for example, the vertices are time ordered: academic papers have
a date or time of publication; species have a time of origination or
speciation.  In food webs vertices are ordered according to tropic level.
(Trophic level, however, is often only an approximate concept and not
precisely defined, which is why some food webs are only approximately
acyclic, containing a few violations of the no-loops condition.)  In
software call graphs, the vertices, representing functions or subroutines,
are ordered according to the software abstraction layer they occupy, and so
forth.

In each of these cases, it is the ordering of the vertices and not their
acyclic structure that is the definitive property of the network.  The
acyclic structure is merely a corollary of the ordering.  In citation
networks, for instance, papers can only cite others that came before them
in time, and this eliminates closed cycles because all paths in the network
must lead backward in time and there are no forward paths available to
close the cycle.  Similarly in food webs species of higher trophic level
prey on those of lower level.  In software graphs functions at higher
levels of abstraction call those at lower levels.  The name ``directed
acyclic graph'' is thus perhaps slightly misleading, focusing our
attention, as it does, on the acyclic property rather than the more
fundamental ordering.  A better name might be ``directed ordered graphs,''
but unfortunately the literature on this topic has long ago settled on the
older name and it seems unwise to try and change it now.

What is important for our purposes, however, is that a sensible random
graph model for these networks should mirror the features seen in the real
world and incorporate an underlying ordering of the vertices that then
drives the acyclic structure.  Thus the correct model is really a ``random
ordered graph'' and this is the approach we take in this
paper~\cite{note1}.

\section{Random directed acyclic graphs with fixed degree sequences}
\label{sec:model}

In this paper we propose two related random graph models of directed
acyclic graphs.  The two models are roughly analogous to the well known
$G(n,m)$ and $G(n,p)$ versions of the standard Poisson random
graph~\cite{ER59}, one fixing the number of edges in the network exactly
and the other fixing only the expected number.  We begin by describing the
``$G(n,m)$'' version, which we introduced previously in
Ref.~\cite{Karrer09a}.  The ``$G(n,p)$'' version, which is introduced for
the first time in this paper, is described in Section~\ref{sec:indepedge}.

Our first model takes as its input an ordered degree sequence consisting of
the in-degree $\kin_i$ and out-degree $\kout_i$ for each vertex $i=1\ldots
n$, where $n$ is the total number of vertices in the network.  The directed
edges in the model are allowed to run only from vertices with higher
indices to vertices with lower, and this constraint enforces the acyclic
nature of the network.  Thus we can have an edge running to vertex~$i$ from
vertex~$j$ only if $i<j$.

Throughout this paper we describe our networks in the language of time
ordering: vertices are ``earlier'' or ``later'' in the network, meaning
they have lower or higher indices, and the vertices with the lowest and
highest indices are referred to as ``first'' and ``last.''  The use of
these terms is purely for convenience and should not be taken as
restricting the model to networks in which the vertices are time ordered.
The concepts we introduce can be applied equally to networks such as food
webs and call graphs in which the ordering has nothing to do with time.

\subsection{Graphical degree sequences}
\label{sec:graphical}

A first important point to notice is that not all degree sequences are
realizable as ordered acyclic graphs of the type described here.  By
analogy with similar issues in other branches of graph theory, we will
refer to realizable degree sequences as \defn{graphical}.

As with all directed graphs, if a degree sequence is to be graphical the
sum of the in-degrees of all vertices must equal the sum of the
out-degrees, since every edge that starts somewhere ends somewhere.  Both
sums are also individually equal to the total number~$m$ of edges in the
network:
\begin{equation}
\sum_{i=1}^n \kin_i = \sum_{i=1}^n \kout_i = m.
\label{eq:totaledges}
\end{equation}
For a directed acyclic graph, however, there are also additional
conditions.  For instance, the first ($i=1$) vertex in the graph can never
have any outgoing edges, since there are no earlier vertices for such edges
to attach to.  Thus $\kout_1=0$ always in a graphical degree sequence.
Similarly $\kin_n=0$.  More generally, we can derive a condition on the
out-degree of every vertex as follows.

It is helpful to visualize in- and out-degrees as sets of ``stubs'' of
edges pointing in and out of each vertex in the appropriate numbers.  To
create a complete network we need to match the stubs in pairs, out with in,
to make whole edges, and a degree sequence is graphical only if all stubs
can be matched while respecting the ordering of the vertices.

The number of stubs outgoing from vertices below vertex~$i$ is
$\sum_{j=1}^{i-1} \kout_j$ and each such stub must be matched with an
ingoing stub at a vertex below~$i$, of which there $\sum_{j=1}^{i-1}
\kin_j$.  The number of ingoing stubs below~$i$ that are left over after we
do this matching is
\begin{equation}
\mu_i = \sum_{j=1}^{i-1} \kin_j - \sum_{j=1}^{i-1} \kout_j.
\label{eq:flux}
\end{equation}
This is the number of ingoing stubs below vertex~$i$ that are available to
attach to outgoing stubs at $i$ and above.  Note that this number is
determined entirely by the degree sequence---it does not depend on any of
the details of which vertices are connected to which others.

Now consider vertex~$i$ itself.  Its out-degree~$\kout_i$ is the number of
its outgoing stubs, and each of those stubs must be matched with an ingoing
one below~$i$.  That means that $\kout_i$ cannot be greater than $\mu_i$
above---if it were, then there would not be enough in-stubs available for
$i$'s out-stubs to attach to and the degree sequence would not be
graphical.  Thus a necessary condition for a degree sequence to be
graphical is
\begin{equation}
\kout_i \le \sum_{j=1}^{i-1} \kin_j - \sum_{j=1}^{i-1} \kout_j.
\label{eq:sums}
\end{equation}
For convenience, we define
\begin{equation}
\lambda_i = \sum_{j=1}^{i-1} \kin_j - \sum_{j=1}^i \kout_j,
\label{eq:lambda}
\end{equation}
so that~\eqref{eq:sums} can be written as
\begin{equation}
\lambda_i \ge 0.
\label{eq:graphical}
\end{equation}
This condition must hold for all~$i$ if the degree sequence is to be
graphical.

Our earlier condition that $\kout_1=0$ trivially implies that
$\lambda_1=0$, and $\kin_n=0$ implies that $\lambda_n=0$ because
\begin{align}
\lambda_n &= \sum_{j=1}^{n-1} \kin_j - \sum_{j=1}^n \kout_j \nonumber\\
  &= (m-\kin_n) - m = 0,
\end{align}
where we have made use of Eq.~\eqref{eq:totaledges}.  Thus we also have
\begin{equation}
\lambda_1 = \lambda_n = 0.
\label{eq:lambda0}
\end{equation}

One might imagine that one could now make a similar argument about the
in-degrees of each vertex and derive a second condition for graphical
sequences of the form:
\begin{equation}
  \sum_{j=i+1}^n \kout_j - \sum_{j=i}^n \kin_j \ge 0.
\end{equation}
This is correct, but in fact it is just another form of the first
condition, Eq.~\eqref{eq:graphical}, as the reader can easily verify by
applying Eq.~\eqref{eq:totaledges}.

Equations~\eqref{eq:graphical} and~\eqref{eq:lambda0} are a necessary
condition for the degree sequence to be graphical.  It's straightforward to
show that they are also sufficient.  The proof is a constructive one: we
build a network starting from the first vertex and working up.
If~\eqref{eq:graphical} holds then at each vertex~$i$ we know that the
number of free in-stubs at lower vertices is at least~$\kout_i$, and hence
there are in-stubs available to attach all of our out-stubs to.  If we
simply choose between the available stubs in any way we like, create the
appropriate edges, and move on to the next vertex, then so long as there
are no unused in-stubs left when we get to the last vertex, which is
guaranteed by Eq.~\eqref{eq:lambda0}, we will have built a complete graph
and hence the sequence is graphical.

Thus Eqs.~\eqref{eq:graphical} and~\eqref{eq:lambda0} are a necessary and
sufficient condition for a graphical degree sequence.

\begin{figure}
\begin{center}
\includegraphics[width=5cm]{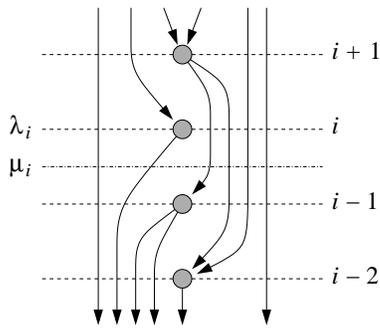}
\end{center}
\caption{The flux $\mu_i$ is equal to the number of edges from vertices $i$
  and above that connect to vertices below~$i$.  The excess flux
  $\lambda_i$ is the number of edges that go around vertex~$i$, connecting
  vertices above to vertices below without passing through~$i$.  In this
  example $\mu_i=6$ and $\lambda_i=5$.}
\label{fig:betalambda}
\end{figure}

The quantities $\mu_i$ and $\lambda_i$ have a simple geometric
interpretation as shown in Fig.~\ref{fig:betalambda}.  If we make a cut in
our graph between vertices $i$ and $i-1$, the quantity~$\mu_i$ is the
number of edges that cross the cut, or the number flowing from higher to
lower vertices.  For this reason, we call $\mu_i$ the \defn{flux} at
vertex~$i$.  (Technically the flux is a property not of the vertex but of
the gap between vertices $i$ and $i-1$, but we have to give it a label so
we choose to label it with the upper of the two vertices.)

The quantity~$\lambda_i$ is equal to the number of edges that flow
``around'' vertex~$i$, meaning the number that run from vertices above~$i$
to vertices below.  We call this quantity the \defn{excess flux} at
vertex~$i$.  Using Eq.~\eqref{eq:flux}, we can show that the flux and
excess flux are related by
\begin{equation}
\mu_i = \lambda_i + \kout_i = \lambda_{i-1} + \kin_{i-1}.
\label{eq:mulambda}
\end{equation}
In the limit of large network size, as we will shortly see, the flux and
excess flux are equal to one another to within a fraction of order $1/n$,
and we will refer to both simply as ``flux'' in this limit.

\begin{figure}
\begin{center}
\includegraphics[width=8cm]{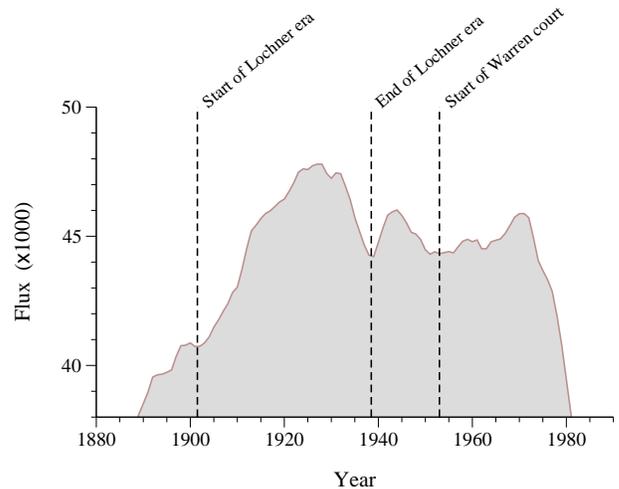}
\end{center}
\caption{Flux~$\mu_i$ for the network of citations between legal opinions
  of the US Supreme Court, plotted as a function of year of publication.
  The three dotted lines highlight dips in the flux and correspond roughly
  to three widely acknowledged shifts in the legal philosophy of the court:
  the start and end of the ``Lochner era,'' during which the court took a
  strong anti-regulatory stance, and the start of the Warren court.  (Note
  that the origin is suppressed on the vertical axis.)}
\label{fig:scotus}
\end{figure}

The flux is a quantity of interest in its own right in real-world networks.
Low values of flux indicate ``bottlenecks'' in a network---lines across
which few edges flow---and high values indicate regions in which there are
many edges.  Figure~\ref{fig:scotus}, for example, shows the measured flux
as a function of time for the network of citations between legal decisions
of the Supreme Court of the United States~\cite{LCSN07}.  A number of dips
in the flux are visible in the figure (marked with dotted lines).  In legal
terms, these dips correspond to temporal divisions between sets of opinions
such that the earlier set is little cited by the later set.  It is a
reasonable guess that these divisions reflect changes in legal thought that
made older opinions obsolete, and indeed each of the three dips highlighted
in the figure corresponds to an acknowledged shift in Supreme Court
jurisprudence, as indicated.

\subsection{Definition of the model}
\label{sec:defmodel}
The definition of our random graph model is now straightforward.  In the
language of ``stubs'' introduced above, a graph on a graphical degree
sequence is created by matching in- and out-going stubs in pairs to create
$m$ complete edges while respecting the ordering of the vertices (meaning
that out-stubs can connect only to earlier in-stubs).  Our model is defined
to be the ensemble of all such matchings in which every matching appears
with equal probability.

This definition is the exact equivalent for directed acyclic graphs of the
standard configuration model for undirected graphs~\cite{MR95}.  In the
configuration model one matches undirected stubs in pairs to create
undirected edges and all matchings appear with equal probability in the
ensemble.  Note that in our model, as in the configuration model,
multiedges are allowed.  That is, the same pair of vertices can be
connected by more than one edge.  (Unlike the configuration model, there
are no self-edges in an acyclic network, since this would violate the
no-cycles rule.)  Multiedges occur in some real-world acyclic networks, but
not in others.  In the model, however, they typically constitute a small
$\Ord(1/n)$ fraction of all edges, and so are negligible in the large
system size limit.  At the same time, a model that admits them is far
easier to study analytically than a model that does not.

Note also that the model includes random ordered trees---which have been
widely studied in the past---as a special case.  If every vertex in the
network (other than the first) has out-degree~1 then the network is
necessarily a tree and the ensemble is uniform over all ordered tree-like
matchings with the given degrees.

Although the model is simple and intuitive, there are---just as with the
configuration model---some subtleties to its definition.  An important
point to notice is that matchings of stubs are not in one-to-one
correspondence with network topologies.  Imagine our stubs to be labeled
somehow, with letters or numbers, so that each one is uniquely
identifiable.  There will then, in general, be many different matchings
that correspond to each possible network topology.  If we take a matching
and simply permute the labels of the out-stubs at a single vertex~$i$, we
produce a new matching corresponding to the same topology.  The number of
distinct such permutations is~$\kout_i!$.  We can similarly permute the
in-stubs at vertex~$i$ for a total of $\kin_i!$ permutations, and the
number of permutations of all stubs at all vertices is then $\prod_i
\kin_i!\kout_i!$.  This, in the simplest case, is the number of matchings
that correspond to each topology.  Since this number is a function solely
of the degree sequence, it is the same for all topologies, and hence if all
matchings occur with equal probability~$p$, then all topologies occur with
equal probability~$p\prod_i \kin_i!\kout_i!$.

\begin{figure}
\begin{center}
\includegraphics[width=8cm]{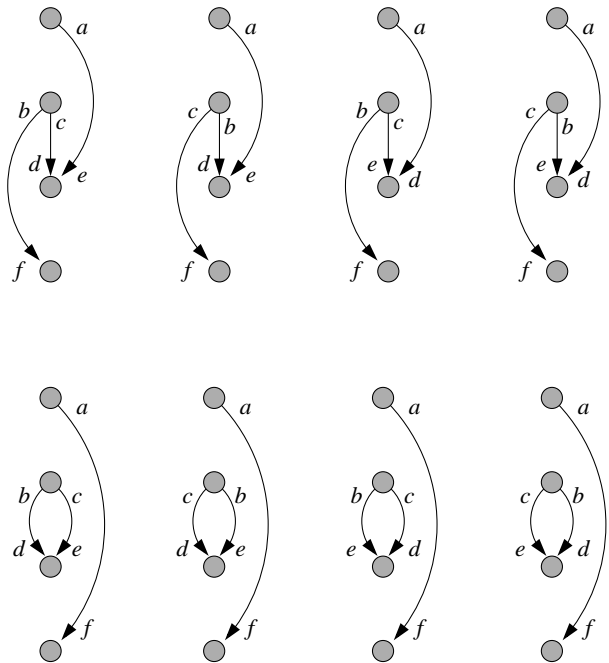}
\end{center}
\caption{Top: a small directed acyclic network with four vertices and three
  edges.  The stubs at each vertex are labeled with letters, and the four
  versions of the graph show the matchings of the stubs generated by
  permuting the stubs at each vertex.  Each permutation generates a
  different matching, so there are in this case four matchings
  corresponding to the same graph, as we would expect since the product
  $\prod_i \kin_i!\kout_i! = 4$ in this case.  Bottom: a second graph with
  the same degree sequence, but now with a multiedge between the two center
  vertices.  There are again four permutations of the stubs as shown, but
  now they correspond to only two different matchings---close inspection
  reveals that the first and fourth matchings are the same, as are the
  second and third.  Thus in this case there are only two matchings for
  this graph.  If all matchings are generated with equal probability, as in
  our model, then the top graph will be generated twice as often as the
  bottom one.}
\label{fig:permutation}
\end{figure}

Unfortunately, there is a complication: if there are multiedges in the
graph then the argument breaks down.  Figure~\ref{fig:permutation} shows
why.  If we identically permute the in-stubs at one end of a multiedge and
the out-stubs at the other end, then we do not generate a new matching---we
get back the same matching we started with.  We see this effect in the
lower half of the figure, where the four distinct permutations of stubs
generate only two distinct matchings.  (The top half of the figure shows
another graph with the same degree sequence but no multiedges and in this
case each permutation generates a unique matching.)

The net result is that our previous calculation overcounts the number of
matchings per topology by a factor of the number of permutations of edges
within multiedges.  If there are no multiedges, then our previous
calculation is correct.  If there are multiedges then the number of
matchings is reduced by a factor of $\prod_{i<j} A_{ij}!$, where $A_{ij}$
is an element of the adjacency matrix, i.e.,~the number of edges between
vertices~$i$ and~$j$.  Since this factor depends on the number and
multiplicity of the multiedges, it follows that in general all topologies
are not sampled with exactly equal probability in our model.

In practice, this is not a significant problem.  The same issue arises in
the configuration model but does not reduce the usefulness of that model.
For the sake of precision, however, we note that although our model samples
matchings with equal probability, it samples topologies with unequal
probabilities that depend on the number and multiplicity of multiedges.

\subsection{Computer generation of networks}

One attractive feature of the model proposed here is that it is
straightforward to generate networks drawn from the model's ensemble on a
computer.  Previous methods for generating directed acyclic graphs have
relied on Monte Carlo techniques~\cite{IC02,MDB00,MWR06} but these methods,
while versatile, are quite slow.  Our model, by contrast, allows us to
generate networks rapidly, in time $\Ord(m)$, where $m$ again is the total
number of edges in the network.  The algorithm, described briefly
in~\cite{Karrer09a}, is based on the scheme outlined in
Section~\ref{sec:graphical} for building a network.  Starting with $n$
vertices and an appropriate number of stubs at each, we go through the
vertices in order from 1 to~$n$.  For each vertex we randomly join its
outgoing stubs to ingoing ones at lower vertices chosen uniformly from the
set of all such in-stubs that are currently unused.  When all stubs have
been matched in this fashion, the network is complete and the algorithm
ends.

It is straightforward to see that indeed this algorithm generates all
matchings with equal probability.  Consider the step of the algorithm at
which out-stubs from vertex~$i$ are matched to suitable in-stubs.  The
number of out-stubs is $\kout_i$ and the number of in-stubs available to
match them to is, by definition, equal to the flux~$\mu_i$.  Thus the
number of different matchings of stubs on this $i$th step is
$N_i=\mu_i!/(\mu_i-\kout_i)!=\mu_i!/\lambda_i!$, where we have used
Eq.~\eqref{eq:mulambda} in the second equality, and the algorithm chooses
between these uniformly at random so that each one occurs with equal
probability $1/N_i$.  Repeating the process for all $n$ vertices generates
a unique matching of the entire graph with probability
\begin{equation}
\prod_{i=2}^n \frac{1}{N_{i}}
  = \prod_{i=2}^n {\lambda_i!\over\mu_i!}.
\label{eq:Nimatching}
\end{equation}
This probability is clearly uniform over all possible matchings since it
depends only on the degree distribution and not on any details of the
matching itself.

The algorithm can be implemented efficiently by maintaining in an ordinary
array a list of currently unclaimed in-stubs from which we choose at random
on every step.  As soon as it is chosen, each stub is erased from the list
by moving the list's last item into its place.  The operations for each
stub can be performed in time $\Ord(1)$, and hence the total running time
is simply proportional to the total number of in-stubs, which is~$m$.

\subsection{Expected number of edges}
\label{sec:edgeprob}

One of the most fundamental properties of our model is the expected number
of directed edges between any two vertices $i$ and $j$.  We will denote
this quantity~$P_{ij}$.  In the limit of large network size $P_{ij}$
becomes small and is equal to the probability that there will be an edge
between $i$ and~$j$.  We assume that $i<j$ in the following calculations,
so that the edge in question always runs from $j$ to~$i$.

\begin{figure}
\begin{center}
\includegraphics[width=5cm]{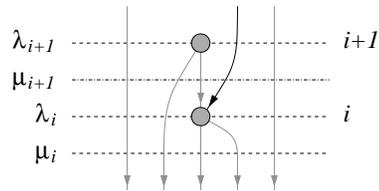}
\end{center}
\caption{The probability that an edge (shown in bold) leaving vertex~$i$
  does not connect to vertex~$i+1$ is given by $\lambda_{i+1}/\mu_{i+1}$.}
\label{fig:flow}
\end{figure}

Consider Fig.~\ref{fig:flow} and consider one of the ingoing edges at
vertex~$i$.  That edge forms part of the flux~$\mu_{i+1}$ immediately
above~$i$ and of that flux $\kout_{i+1}$ edges, chosen uniformly at random,
originate at vertex~$i+1$, while the remaining
$\mu_{i+1}-\kout_{i+1}=\lambda_{i+1}$ flow around~$i+1$, forming the excess
flux at $i+1$.  The probability that our particular edge is one of the ones
flowing around $i+1$, i.e., that it does not originate at vertex~$i+1$, is
thus simply $\lambda_{i+1}/\mu_{i+1}$.

If our edge is to originate at vertex~$j$, it must flow in this way around
every intervening vertex from $i+1$ all the way up to~$j-1$, and then
finally it must originate at vertex~$j$, which it does with probability
$\kout_j/\mu_j$.  Multiplying the probabilities together, we find that the
total probability of this particular edge originating at vertex~$j$ is
\begin{equation}
{\kout_j\over\mu_{j}} \prod_{l=i+1}^{j-1} {\lambda_l\over\mu_{l}}
  = \kout_j {\prod_{i+1}^{j-1} \lambda_l \over \prod_{i+1}^j \mu_l}.
\end{equation}
This is just for one of the ingoing edges at vertex~$i$.  There are
$\kin_i$ such edges in all, so the total expected number of edges from $j$
to~$i$ is
\begin{equation}
P_{ij} = \kin_i \kout_j
         {\prod_{i+1}^{j-1} \lambda_l \over \prod_{i+1}^j \mu_l}.
\label{eq:pij1}
\end{equation}

We will find it convenient to write this expression in the form
\begin{equation}
P_{ij} = {\kin_i\kout_j\over m} f_{ij},
\label{eq:pij2}
\end{equation}
where
\begin{equation}
f_{ij} = m {\prod_{i+1}^{j-1} \lambda_l \over \prod_{i+1}^j \mu_l}.
\label{eq:fij1}
\end{equation}
The quantity $\kin_i\kout_j/m$ is the expected number of edges between $i$
and $j$ in an ordinary (not acyclic) directed random graph with the same
degree sequence, so $f_{ij}$ represents the factor by which that expected
number is modified in the acyclic graph.  Alternatively, $f_{ij}$~is $m$
times the probability that a single in-stub at vertex~$i$ is connected to a
single out-stub at vertex~$j$.  (The probability itself vanishes in the
limit of large graph size but with the inclusion of the factor of~$m$ we
get a quantity that tends to a nonzero limit, which will be useful when we
come to consider properties of the graph as $n\to\infty$.)

One complication in the expression for $f_{ij}$ occurs if any flux in the
denominator is zero.  The expression gives the correct answer of zero for
$P_{ij}$ if we adopt the convention that $0/0=1$.  However, it's usually
better to analyze a graph divided by a zero flux cut as two independent
graphs, since no edges cross the cut in such a network and the network
forms two separate components.  A network with zero \emph{excess} flux does
not necessarily form two separate components---the two parts of the network
can by joined by a single common vertex at the top of one part and the
bottom of the other---but the two parts can be treated independently
anyway, with the shared vertex, if any, participating in both parts.
Hence, in the following, we assume that $\mu_i \neq 0$ and $\lambda_i \neq
0$ except for $i=1$ and $i=n$.

Another useful expression for~$f_{ij}$ can be derived by multiplying both
sides of Eq.~\eqref{eq:fij1} by $f_{i'j'}$ with the condition that $i$
and~$i'$ are both less than $j$ and~$j'$.  Then
\begin{align}
f_{ij} f_{i'j'} &= {\prod_{l=i+1}^{j-1} \lambda_l\over
                    \prod_{l=i+1}^j \mu_l} \,
                   {\prod_{l=i'+1}^{j'-1} \lambda_l\over
                    \prod_{l=i'+1}^{j'} \mu_l}\
                 = {\prod_{l=i+1}^{j'-1} \lambda_l\over
                    \prod_{l=i+1}^{j'} \mu_l} \,
                   {\prod_{l=i'+1}^{j-1} \lambda_l\over
                    \prod_{l=i'+1}^{j} \mu_l} \nonumber\\
                &= f_{ij'} f_{i'j}.
\end{align}
Thus we can freely swap indices on a product of two overlapping~$f$s.  In
particular, if we set $i'=1$ and $j'=n$, we find that
\begin{equation}
f_{ij} = {f_{in}f_{1j}\over f_{1n}},
\end{equation}
and $f_{ij}$ thus factors into a product of independent functions of $i$
and~$j$.  This result is of some practical use, since it implies that in
order to calculate $f_{ij}$ or $P_{ij}$ for any $i$ and~$j$ we need only
the quantities $f_{in}$ and $f_{1j}$, which are $\Ord(n)$ in number and
take~$\Ord(n)$ time to calculate.  Once these are known, we can calculate
any $P_{ij}$ in $\Ord(1)$ time, which is as fast as the corresponding
calculation for the configuration model, and far faster than direct
application of Eq.~\eqref{eq:pij1}, which takes $\Ord(n)$ time on average
for each~$P_{ij}$.

Perhaps the simplest way to implement this idea in practice is to define
the two ``dimensionless'' quantities
\begin{equation}
a_i = {f_{in}\over f_{1n}},\qquad
b_j = {f_{1j}\over f_{1n}},
\label{eq:defsab}
\end{equation}
so that
\begin{equation}
f_{ij} = f_{1n} a_i b_j.
\label{eq:fab}
\end{equation}
Clearly $a_1 = b_n = 1$ and, substituting from Eq.~\eqref{eq:fij1} into
Eq.~\eqref{eq:defsab}, we find the values for other $i,j$ to be
\begin{subequations}
\begin{align}
\label{eq:formai}
a_i &= {\prod_{l=2}^i \mu_l \over \prod_{l=2}^i\lambda_l}
     = \prod_{l=2}^i \biggl( 1 + {\kout_l\over\lambda_l} \biggr),\\
\label{eq:formbj}
b_j &= {\prod_{l=j+1}^n \mu_l \over \prod_{l=j}^{n-1} \lambda_l}
     = \prod_{l=j}^{n-1} \biggl( 1 + {\kin_l\over\lambda_l} \biggr),
\end{align}
\label{eq:formab}
\end{subequations}
where we have made use of Eq.~\eqref{eq:mulambda}~\cite{note2}.  We will
use these expressions in a number of calculations in the following
sections.

\subsection{Assortativity}
\label{sec:assort}
As an example of the application of the calculations in the previous
section, consider vertex correlations or ``assortativity'' in acyclic
networks~\cite{Newman03c}.

Consider a quantity $x$ defined on all vertices~$i$ of a network.  The
network is said to be assortative with respect to~$x$ if edges tend to
connect vertices with similar values of~$x$, high with high and low with
low.  Conversely, if edges connect dissimilar values, high with low and
\textit{vice versa}, then the network is said to be disassortative.
Assortativity can be quantified by calculating a standard Pearson
correlation coefficient~$r$ over all pairs of values~$x_i,x_j$ on vertices
$i,j$ connected by an edge.  Positive values of~$r$ indicate assortative
networks, negative values disassortative ones.

In a directed network, such as the acyclic networks considered here, more
complex types of correlations are also possible.  For instance, one can
consider two different quantities, $x$~and~$y$, each defined on all
vertices, and then ask about the correlations between pairs of values
$x_i,y_j$ on vertices $i,j$ connected by a directed edge from~$j$ to~$i$.
(The simpler example above with only one quantity~$x$ can be considered as
the special case in which $y=x$.)  Again one can calculate a correlation
coefficient that quantifies the level of assortativity or disassortativity.
The correlation coefficient is given explicitly in terms of the standard
adjacency matrix by
\begin{equation}
r = {1\over\sigma_X\sigma_Y}\,
    \biggl[ {1\over m} \sum_{ij} A_{ij} x_i y_j -
    \mu_\mathrm{in}\mu_\mathrm{out} \biggr],
\label{eq:corrcoefficient}
\end{equation}
where
\begin{equation}
\mu_\mathrm{in} = {1\over m} \sum_i \kin_i x_i,\qquad
\mu_\mathrm{out} = {1\over m} \sum_j \kout_j x_j,
\end{equation}
and
\begin{subequations}
\begin{align}
\sigma_X^2 &= {1\over m} \sum_i \kin_i x_i^2
            - \mu_\mathrm{in}^2, \\
\sigma_Y^2 &= {1\over m} \sum_j \kout_j y_j^2
            - \mu_\mathrm{out}^2.
\end{align}
\end{subequations}

Conventional random graph models such as the configuration model show no
assortativity with respect to any quantity~$x$, but random acyclic graphs
can have nonzero assortativity.  Consider Eq.~\eqref{eq:corrcoefficient}
for the acyclic case and notice that the only dependence on~$A_{ij}$ is in
the first term of the numerator.  All the other terms depend only on the
degree sequence of the network, and hence are constant for our acyclic
graph model over all members of the model ensemble.  Averaging over the
ensemble and noting that the model average of~$A_{ij}$ is simply~$P_{ij}$
from Eq.~\eqref{eq:pij2}, we find that within our model
\begin{align}
r &= {1\over\sigma_X\sigma_Y}\,
    \biggl[ {1\over m} \sum_{ij} P_{ij} x_i y_j -
    \mu_\mathrm{in}\mu_\mathrm{out} \biggr] \nonumber\\
  &= {1\over\sigma_X\sigma_Y}\,
    \biggl[ {f_{1n}\over m^2} \sum_{i<j} a_i b_j \kin_i \kout_j x_i y_j -
    \mu_\mathrm{in}\mu_\mathrm{out} \biggr],
\end{align}
where we have used Eq.~\eqref{eq:fab}.  In general, this expression can
give nonzero values of~$r$.  We will see some examples in
Section~\ref{sec:empirical} for the particular case of assortativity with
respect to vertex degree~\cite{PVV01,Newman02f,MSZ04}, such as the case in
which $x_i=\kin_i$ and $y_j=\kout_j$.

\subsection{Large system-size limit}
\label{sec:continuum}

The developments so far are for a network of finite size with a specified
degree sequence.  Like other random graph models, however, random acyclic
graphs become significantly simpler in a number of ways in the limit of
large graph size.  We examine that limit in this section.

Let the number of vertices in our network be~$n$ as previously.  In the
limit of large~$n$ we can no longer specify the complete degree sequence,
since there are an infinite number of vertices, so, as with other random
graphs, we specify instead a degree distribution, which is a joint
probability distribution over in- and out-degrees as a function of vertex
order.  We define a ``time'' variable $t=i/n$ for the $i$th vertex, which
falls in the range $0<t\le1$, then let $p_t(\kin,\kout)$ be the probability
that a vertex at time~$t$ has in- and out-degrees $\kin$ and~$\kout$.
Since vertices are uniformly distributed in time, this distribution is
related to the overall (joint) degree distribution of the network by a
simple integral:
\begin{equation}
p(\kin,\kout) = \int_0^1 p_t(\kin,\kout) \>\d t.
\end{equation}

Unfortunately the full distribution $p_t(\kin,\kout)$ is usually impossible
to measure for an observed network: measuring it would require us to build
a double histogram of $\kin$ and~$\kout$ for many small intervals of~$t$
and none of the real-world networks we have examined are large enough to
give acceptable statistics for such a histogram.  Luckily, however, it
turns out that many interesting characteristics of the network can be
calculated with a knowledge only of the moments of the degree distribution,
and in most cases only the first moment, i.e.,~the mean degree.

The mean in- and out-degrees at time~$t$ are given by
\begin{align}
\kinav(t) &= \sum_{\kin=0}^\infty\,\sum_{\kout=0}^\infty
             \kin p_t(\kin,\kout), \nonumber\\
\koutav(t) &= \sum_{\kin=0}^\infty\,\sum_{\kout=0}^\infty
              \kout p_t(\kin,\kout),
\end{align}
and the overall average degree~$c$ of the network is
\begin{equation}
c = \int_0^1 \kinav(t) \>\d t = \int_0^1 \koutav(t) \>\d t.
\label{eq:knorm}
\end{equation}
Both $\kinav(t)$ and $\koutav(t)$ are easily measured in practice (at least
approximately) by performing running averages of the observed degrees over
suitably chosen time intervals.

For many of the calculations presented here we will use the rescaled
quantities
\begin{equation}
\kappain(t) = {\kinav(t)\over c},\qquad
\kappaout(t) = {\koutav(t)\over c},
\end{equation}
which satisfy the normalization conditions
\begin{equation}
\int_0^1 \kappain(t) \>\d t = \int_0^1 \kappaout(t) \>\d t = 1.
\label{eq:kappanorm}
\end{equation}
The quantity $\kappain(t)\>\d t$ is the fraction of all in-stubs that are
attached to vertices in the range $t$ to $t+\d t$, and similarly
for~$\kappaout(t)\>\d t$.  The \emph{numbers} of stubs are given by
$m\kappain(t)\>\d t$ and $m\kappaout(t)\>\d t$, since $m$ is the total
number of stubs of each kind in the whole network.

The flux below vertex~$i$ in the network is given by integrating these
quantities up to a given vertex thus:
\begin{equation}
\mu_i = m \int_0^t \bigl[ \kappain(t') - \kappaout(t') \bigr] \d t'.
\label{eq:largeflux}
\end{equation}
where $t=i/n$ as before.  Note that, assuming the degree distribution
remains constant as the network becomes large, the integral for given~$t$
also remains constant, but $m=nc$ grows with network size.  Thus the flux
becomes arbitrarily large as $n\to\infty$.  For our purposes it is better
to use a quantity that remains constant as $n$ varies and so we define a
rescaled flux
\begin{equation}
\mu(t) = {\mu_i\over m}
       = \int_0^t \bigl[ \kappain(t') - \kappaout(t') \bigr] \d t'.
\end{equation}
In the large system size limit, there is no difference between the
flux~$\mu$ and the excess flux~$\lambda$: the two differ only by the number
of stubs at a single vertex, which is a vanishing fraction of~$m$ in the
limit of large network size, and hence $\lambda_i$ also varies as~$m$ and
the rescaled excess flux $\lambda(t)=\lambda_i/m$ is given by
\begin{equation}
\lambda(t) = \int_0^t \bigl[ \kappain(t') - \kappaout(t') \bigr] \d t'.
\end{equation}
Physically $\mu(t)$ and $\lambda(t)$ are both equal to the fraction of
edges that run from vertices after~$t$ to vertices before.

Applying these definitions, we can now calculate a variety of quantities
in the $n\to\infty$ limit.  To calculate the probability of connection
between two vertices, we start with Eq.~\eqref{eq:formai}:
\begin{equation}
a_i = \prod_{l=2}^i \biggl( 1 + {\kout_l\over\lambda_l} \biggr)
    = \exp \Biggl[ \sum_{l=2}^i \ln\biggl(1 + {\kout_l\over\lambda_l}
           \biggr) \Biggr].
\end{equation}
Observing, as above, that $\lambda_l$ goes as $m$ in the large system size
limit while $\kout_l$ remains constant and keeping terms to leading order,
this becomes
\begin{equation}
a_i = \exp \Biggl[ \sum_{l=2}^i {\kout_l\over\lambda_l} \Biggr].
\end{equation}
And in the limit of large~$n$, the sum becomes an integral:
\begin{equation}
a(t) = \exp \biggl[ \int_0^t {\kappaout(t')\over\lambda(t')} \>\d t' \biggr].
\end{equation}
Similarly, defining $u=j/n$, Eq.~\eqref{eq:formbj} becomes
\begin{equation}
b(u) = \exp \biggl[ \int_u^1 {\kappain(u')\over\lambda(u')} \>\d u' \biggr],
\end{equation}
and substituting both into Eq.~\eqref{eq:fab} we get
\begin{equation}
f(t,u) = f(0,1)\,a(t)\,b(u),
\label{eq:ftu}
\end{equation}
where $f_{ij} = f(i/n,j/n)$.  Physically, $f(t,u)$ is $m$ times the
probability that an in-stub at time~$t$ is connected to an out-stub at
time~$u$.  The normalizing constant $f(0,1)$ can be calculated by noting
that every in-stub must be connected to \emph{some} out-stub, which means
that
\begin{equation}
\int_t^1 f(t,u)\,\kappaout(u) \>\d u = 1.
\end{equation}
Substituting for $f(t,u)$ from Eq.~\eqref{eq:ftu} and setting $t=0$ then
gives
\begin{equation}
f(0,1) = \biggl[ \int_0^1 b(u) \kappaout(u) \>\d u \biggr]^{-1},
\label{eq:f01}
\end{equation}
where we have made use of $a(0)=1$.  If we instead normalize by integrating
over~$t$ we get the alternative form
\begin{equation}
f(0,1) = \biggl[ \int_0^1 a(t) \kappain(t) \>\d t \biggr]^{-1},
\end{equation}
which gives the same answer but may be more convenient in some cases,
depending on the forms of $\kappain$ and~$\kappaout$.

Armed with a value for $f(t,u)$ we can now calculate the expected number of
edges between two vertices in the network from Eq.~\eqref{eq:pij2}:
\begin{equation}
P_{ij} = {\kin_i\kout_j\over m} f(i/n,j/n).
\end{equation}
Alternatively, we can average this expression over the distributions of
$\kin$ and $\kout$ to get the average number of edges between a vertex
at~$t$ and another at~$u$:
\begin{equation}
P(t,u) = {\kinav(t)\koutav(u)\over m} f(t,u)
       = {c\over n} \kappain(t)\kappaout(u)f(t,u).
\end{equation}
Since $f(t,u)$ is independent of~$n$ for given $\kappain(t)$
and~$\kappaout(u)$, $P_{ij}$~[and $P(t,u)$] goes as $1/n$ in a sparse graph
as graph size becomes large and hence vanishes in the limit.  This allows
us to interpret $P_{ij}$ as a probability of connection between vertices in
the $n\to\infty$ limit---the expected number of edges and the probability
of connection are the same when both become small.

We also note in passing the following useful relation between $\lambda(t)$
and $f(t,u)$.  From Eq.~\eqref{eq:pij1} we have
\begin{equation}
P_{i-1,i} = {\kin_{i-1}\kout_i\over\mu_i},
\end{equation}
so that $f_{i-1,i} = m/\mu_i$.  Setting $t=i/n$ as before and
$\mu_i/m=\lambda(t)$, this implies that
\begin{equation}
\lambda(t) = {1\over f(t,t)}.
\end{equation}

\subsection{Examples}
To illustrate the application of these results let us look at some concrete
examples.  Consider a network with average degrees $\kinav(t)=2c(1-t)$ and
$\koutav(t)=2ct$, where $c$ is now a free parameter controlling the overall
mean degree.  Then
\begin{equation}
\kappain(t) = 2(1-t),\qquad \kappaout(u) = 2u,
\label{eq:cascade}
\end{equation}
and we find that
\begin{equation}
f(t,u) = \frac{1}{2(1-t)u}
\end{equation}
and
\begin{equation}
P(t,u) = \frac{2c(1-t) \times 2cu}{2m(1-t)u} = \frac{2c}{n},
\end{equation}
where we have used $m=nc$ in the second equality.

Thus the expected number of edges between every pair of vertices in this
case is the same, and indeed one could exploit this fact to create a
network with the degree sequence above by taking an initially empty graph
and placing a directed edge between each vertex pair with uniform
probability $2c/n$, oriented to point from the ``later'' vertex to the
``earlier'' one.  Such a model has been studied previously as a model of
food webs, in which context it is known as the \defn{cascade
  model}~\cite{CN85}.  It's easy to see that the cascade model produces
networks with a given degree sequence uniformly at random and thus is
approximately equivalent to an acyclic random graph with the same degree
sequence as described in this paper.  The equivalence is only approximate:
the cascade model has a Bernoulli distribution of edges between any two
vertices while our model has a Poisson distribution.  This difference,
however, vanishes in the limit of large graph size, where the edge
probability becomes small, and thus in this limit the two models are the
same.

More generally, consider a model where a Poisson distributed number of
directed edges is placed between all pairs of vertices $i,j$ with $i<j$.
If the mean of the Poisson distribution for each vertex pair can be written
as a product of a quantity~$r_i$ that depends on $i$ but not on~$j$ and a
quantity~$s_j$ that depends on~$j$ but not on~$i$, then the model produces
acyclic random graphs conditioned on the degree sequence.  To prove this we
write the probability~$P$ of generating a particular graph thus:
\begin{equation}
P = \prod_{i<j} \e^{-s_i r_j}\frac{(s_i r_j)^{A_{ij}}}{A_{ij}!}
  = \frac{\prod_{i<j} \e^{-s_i r_j}}{\prod_{i<j}A_{ij}!}
    \prod_i s_i^{\kin_i} r_i^{\kout_i}.
\end{equation}
The factor $\prod_{i<j} \e^{-s_i r_j}$ is a constant for all graphs and the
factor $\prod_i s_i^{\kin_i} r_i^{\kout_i}$ is constant for a given degree
sequence.  Thus the only variation in the probability~$P$ for graphs of
given degree sequence comes from the factor $\prod_{i<j} A_{ij}!$.  But
this is the same factor by which the probability of such graphs varies in
the random acyclic graph model---see Section~\ref{sec:defmodel}---and thus,
for a given degree sequence, the model above produces graphs with the same
probabilities as the random acyclic graph and the two models have identical
ensembles.  The cascade model is a particularly simple instance of this
situation in which $r_i$ and $s_j$ are both constant.
 
As another example, we consider networks with power-law degree
distributions, which have received a lot of attention in the recent
networks literature.  In particular, for reasons that will shortly become
clear, we consider networks generated by linear preferential attachment
processes~\cite{BA99b}, which naturally generate directed acyclic graphs
and have long been used as models of citation networks~\cite{Price76}.  We
consider the general model in which vertices added continually to a growing
network make $c$ directed connections each to previously existing vertices
chosen at random in proportion to the current in-degrees of those vertices
plus a constant~$r$.  This process produces networks with overall in-degree
distributions having a power-law tail $p(k) \sim k^{-\alpha}$ where
$\alpha=2+r/c$~\cite{Price76,DMS00}.  In the notation used in this paper
the average in-degree as a function of time is given by~\cite{DMS00}:
\begin{equation}
\kappain(t)= (\alpha-2) (t^{-1/(\alpha-1)}-1),
\end{equation}
and $\kappaout(u)=1$.

Let us consider a random directed acyclic graph built on degree sequences
generated by the linear preferential attachment model and let us calculate
the probability of connection between vertices.  Feeding the expressions
above for $\kappain(t)$ and $\kappaout(u)$ into our earlier formulas, we
find that
\begin{equation}
f(t,u) = \frac{1}{(\alpha-1)(1-t^{1/(\alpha-1)})u^{(\alpha-2)/(\alpha-1)}}
\end{equation}
and
\begin{align}
P(t,u) 
       = c(\alpha-2)i^{-1/(\alpha-1)}j^{-(\alpha-2)/(\alpha-1)},
\end{align}
where again $t=i/n$ and $u=j/n$.  Remarkably, this is precisely the average
probability of an edge between vertices in the preferential attachment
model itself~\cite{DM02}.  Indeed, as we will shortly show, the linear
preferential attachment ensemble and the ensemble of the random acyclic
graph with the same degree sequence are actually identical, because linear
preferential attachment, conditioned on the degree sequence, produces
matchings uniformly at random, which is precisely the condition for the
random acyclic graph.  Thus, not only is $P(t,u)$ the same for the two
models, but all properties of the models are identical and one can properly
say that the linear preferential attachment model is a special case of the
random directed acyclic graph.

This is an important point.  It is often claimed that networks produced by
the linear preferential attachment process are, in some sense, not really
random, being nonuniform in their ensemble properties because they are
grown according to a nonequilibrium growth process.  In fact, however, this
is not the case.  Once the acyclic nature of the networks is taken into
account, the ensemble of the linear preferential attachment model is
perfectly uniform for a given degree sequence.

To prove this we compute the probability of a particular matching being
produced by the linear preferential attachment model as a function of
in-degree sequence.  An outgoing edge at a newly added vertex~$j$ in the
growing preferential attachment network attaches to a previous vertex~$i$
with probability proportional to $i$'s current in-degree~$\kin_i$ plus the
constant~$r$.  The correctly normalized probability of attachment is
\begin{equation}
{\kin_i+r\over\sum_{i=1}^{j-1} (\kin_i+r)}
  = {\kin_i+r\over m+(j-1)r},
\label{eq:edgeprob}
\end{equation}
where $m=\sum_i \kin_i$ is the current number of edges in the network.  The
probability of the entire matching is given by the product of this
expression over all edges.  Let us consider the numerator and denominator
of the product separately, starting with the numerator.

The current in-degree of vertex~$i$ is 0 when the first edge attaches to
it, 1~when the second edge attaches, and so forth.  Hence the factors for
vertex~$i$ in the numerator are
\begin{equation}
r(1+r)\ldots(\kin_i-1+r) = {\Gamma(\kin_i+r)\over\Gamma(r)},
\end{equation}
where $\kin_i$ now represents the final in-degree of~$i$ at the end of the
growth process and $\Gamma(x)$ is the standard gamma function.  Taking the
product over all vertices, the complete numerator is $\prod_{i=1}^{n-1}
\Gamma(\kin_i+r)/\Gamma(r)$.  (There is no term for the last vertex since
it necessarily has no ingoing edges.)

For the denominator, we note that the number of edges~$m$ in the network
increases by one for each edge added and takes the value $(j-2)c$ for the
first edge added with vertex~$j$ and $(j-1)c-1$ for the last.  Thus the
factors in the denominator corresponding to the edges added with vertex~$j$
give
\begin{align}
& [(j-2)c+(j-1)r]\ldots[(j-1)c-1+(j-1)r] \nonumber\\
& \hspace{10em} {} = {\Gamma((j-1)(c+r))\over\Gamma((j-1)(c+r)-c)},
\end{align}
and the complete denominator is
\begin{equation}
\prod_{j=2}^n {\Gamma((j-1)(c+r))\over\Gamma((j-1)(c+r)-c)}
  = \prod_{i=1}^{n-1} {\Gamma(i(c+r))\over\Gamma(i(c+r)-c)}.
\end{equation}

Dividing numerator by denominator, the complete probability for the
matching is then
\begin{equation}
P = \prod_{i=1}^{n-1} {\Gamma(\kin_i+r)\over\Gamma(r)}\,
    {\Gamma(i(c+r)-c)\over\Gamma(i(c+r))}.
\end{equation}
Since this probability depends only on the degree sequence and not on any
details of which vertices attach to which others, it follows that the
preferential attachment process generates all matchings with a given degree
sequence with the same probability, and hence that the set of networks with
that degree sequence constitutes a random directed acyclic graph of the
type considered in this paper.

Note that a calculation similar to the one above can be performed for a
model in which out-degree is not the same for every vertex, but varies from
one vertex to another, or a network in which the parameter~$r$ varies
between vertices.  The probability of a particular matching for such a
model is still a function only of the degrees and other parameters and not
of the pattern of connections in the network and hence the network is still
a random graph of the type considered here.

\section{Random directed acyclic graphs with independent edge probabilities}
\label{sec:indepedge}

In this section we define the second of our two random graph models for
acyclic graphs.  In this model rather than fixing the degree of each vertex
we fix only the expected degree.  As discussed in the introduction the
model is in some ways analogous to the $G(n,p)$ model of Erd\H{o}s and
R\'enyi~\cite{ER59} for ordinary (Poisson) random graphs, while the
previous model is the equivalent of~$G(n,m)$.

We have seen that it is possible in our previous model to calculate the
probability of an edge between any pair of vertices.  However, in that
model edges are not independent because the presence of one edge connecting
to a given vertex~$i$ reduces the number of stubs available for other edges
and hence reduces the probability of edges from other vertices.  In the
limit of large network size, the probabilities for edges to and
from intervals $\d t$ and $\d u$ become independent, but even in this limit
edges that share the same exact vertex, either as source or target, remain
correlated.

The same phenomenon is also seen in other random graph models, such as the
configuration model, in which degrees are also fixed and the presence of
one edge to a vertex reduces the probability of others.  In that case,
researchers have found it useful to study a slightly different model in
which edges are placed with the same probability as in the configuration
model, but independently~\cite{CL02a,CL02b,BJR07}.  The same strategy turns
out also to work well in the case of acyclic graphs.  The resulting model
is described in this section.

\subsection{Definition of the model}
Our second model is defined as follows: starting with an empty graph of $n$
vertices we generate for each pair of vertices $i,j$, with $i<j$, a Poisson
distributed number with mean $P_{ij}$ and place that number of edges
between $i$ and~$j$, pointing from $j$ to~$i$.  The values of $P_{ij}$ are
typically calculated from a desired degree sequence using
Eq.~\eqref{eq:pij2}, and the resulting network trivially has the same
expected number of edges between every vertex pair as the network generated
by our first model with the same degree sequence, but the edges are now, by
construction, independent.

Since the number of edges between every vertex pair is Poisson distributed,
so also is the total number of edges~$m$.  Thus an equivalent way to create
networks drawn from this model is to generate a Poisson distributed random
number~$m$ with mean equal to the desired expected number of edges, then
distribute those edges at random over the graph in proportion to~$P_{ij}$.
This second method for generating networks is a more efficient one for
numerical work but the first is more convenient for analytic treatment of
the model.

The principal disadvantage of this model is that it does not allow us to
fix the exact degrees of each vertex.  Instead we can only fix the expected
degrees~$\kinav_i$ and~$\koutav_i$.  The expected in-degree, for instance,
is given by $\sum_{j=i+1}^n P_{ij}$, which is by definition equal to the
value of $\kin_i$ used to calculate $P_{ij}$ in the first place.  In other
words, the network has expected degrees equal to the chosen degree
sequence, but the actual degrees may be different.

In fact, since the numbers of edges are Poisson independent variables, the
in-degree will also be Poisson distributed with mean~$\kin_i$ (and
similarly for the out-degree).  Note however that this does not mean that
the overall distribution of the degrees at any time has to be Poisson,
since the distribution from which the means themselves are drawn can be
anything we like and the overall distribution of degrees is a convolution
of this distribution and the Poisson distribution.

The expected degrees also need not be integers, so this model allows a
slight generalization of the previous one in that the values of $\kin_i$
and $\kout_i$ we use to calculate $P_{ij}$ need not be integers.  Indeed we
could generalize the model considerably further, since in principle we can
choose the values of the $P_{ij}$ to be anything we want, including values
that cannot be generated from Eq.~\eqref{eq:pij2} by any choice of degrees.
Any values, for example, that do not take the product form of
Eq.~\eqref{eq:pij2} fall in this category.  In this paper, however, we will
mostly be concerned with choices of $P_{ij}$ that correspond to an
underlying choice of expected degrees.

\subsection{Computer generation of networks}

It is less straightforward to numerically generate networks drawn from the
ensemble of our second model than of our first.  The basic approach is as
outlined above: given the expected degrees, we calculate the expected
number of edges by summing $\overline{m}=\sum_{i=1}^n \kin_i$ and then
generate a Poisson distributed number with this mean, which will be the
actual number of edges~$m$.

To place these $m$ edges with the appropriate probabilities we need to be
able to randomly generate vertex pairs with probabilities proportional
to~$P_{ij}$.  This can conveniently be achieved by making use of the
product form~\eqref{eq:pij2} of~$P_{ij}$.  We draw a value for $i$ from the
marginal probability distribution, which goes as $\sum_{j=i+1}^n
P_{ij}=\kin_i$, using a standard transformation method, which takes
$\Ord(\log n)$ time.  Then we draw a value for $j$ between $i+1$ and $n$ in
proportion to $\kout_j b_j$, again using the transformation method.  Then
we place an edge between $i$ and~$j$ and repeat for the next edge.  When
all $m$ edges have been placed the graph is complete.  The whole process
takes $\Ord(n)$ time for set-up and $\Ord(m\log n)$ for selection and
placing of edges, or $\Ord(n+m\log n)$ time in total, which is $\Ord(n\log
n)$ on a graph with fixed degree distribution so that $m\propto n$.

\section{Comparison with empirical data}
\label{sec:empirical}
Our expressions for edge probabilities allow us to make a comparison
between our model networks and their counterparts in the real world.  We
focus on citation networks, which are the largest and best documented
examples of acyclic networks.

The simplest comparison we could make would be a direct comparison of edge
probabilities~$P_{ij}$.  However, the value of $P_{ij}$ is strongly
influenced by the degrees of vertices---the initial factor of
$\kin_i\kout_j$ in Eq.~\eqref{eq:pij2}---which makes comparison plots noisy
and difficult to interpret by eye.  A cleaner comparison is of the stub
probability~$f_{ij}$, Eq.~\eqref{eq:fij1}, which is $m$ times the
probability that a stub at vertex~$i$ is connected to a stub at vertex~$j$.

We can make an estimate of $f_{ij}$ for an observed network by taking a
window of vertices around~$i$ and another around~$j$, counting the number
of edges between vertices in the two windows, and then dividing in turn by
the number of in-stubs in the first window and out-stubs in the second and
multiplying by $m$~\cite{note3}.  If the windows are large enough to
provide good statistics but small enough to span only a relatively narrow
range of $i$ and~$j$ then one can get good estimates of the mean stub
probability this way.

\begin{figure*}
\begin{center}
\includegraphics[width=14cm]{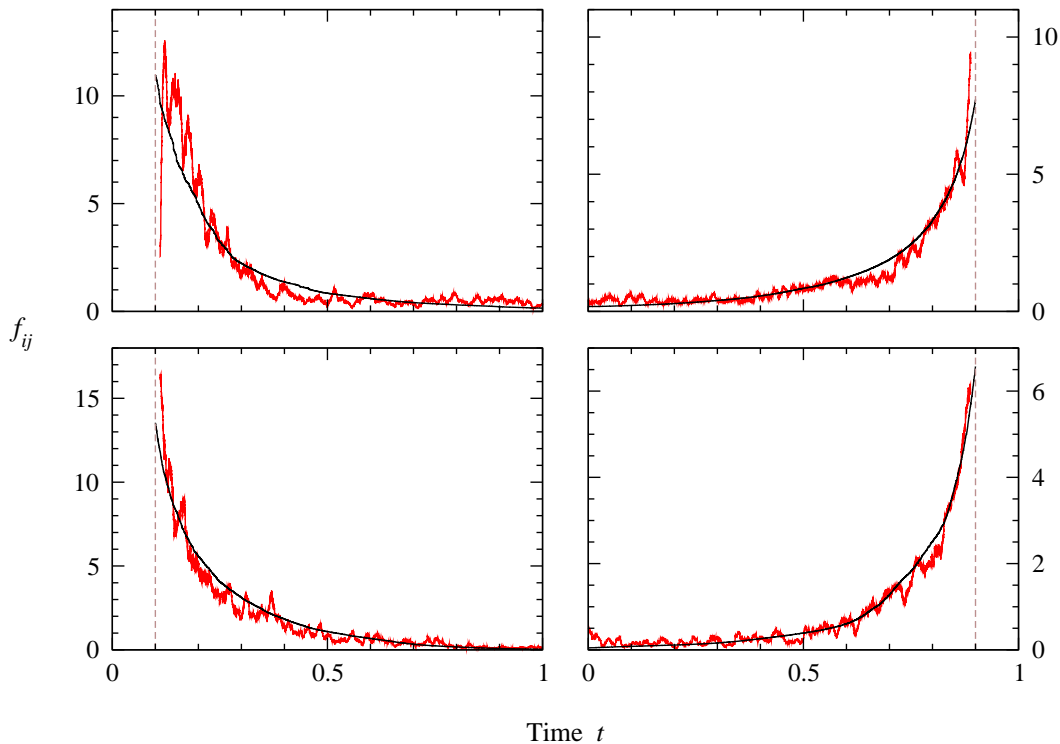}
\end{center}
\caption{Comparison of empirical measurements (red) and analytic
  predictions (black) of $f_{ij}$ for the two citation networks described
  in the text: preprints on high-energy physics (top) and cases of the
  United States Supreme Court (bottom).  The left panel in each case shows
  $f_{ij}$~for citations from times~$t$ to time~$0.1$ (indicated by dashed
  line).  The right panel shows $f_{ij}$~for citations to times~$t$ from
  time~$0.9$.  Empirical measurements were averaged over windows of size
  $300$ vertices.}
\label{fig:hepthf}
\end{figure*}

In Fig.~\ref{fig:hepthf} we show the results of such measurements for two
citation networks.  The first is a network of citations between academic
papers in the area of theoretical high-energy physics, which we studied
previously in Ref.~\cite{Karrer09a}.  This data set comprises $27\,221$
papers posted in the ``hep-th'' section of the Physics E-print Archive at
arxiv.org between January 1992 and February 2003.  The data set was
compiled by the organizers of the KDD Cup challenge, a data analysis
competition run as part of the annual ACM SIGKDD conference, and
incorporates citations extracted from data held in the SPIRES database at
the Stanford Linear Accelerator Center.

The second data set is a network of citations between $26\,084$ legal
decisions handed down by the United States Supreme Court, from the time of
the court's inception in 1789 until 2006, as compiled by
Leicht~\etal~\cite{LCSN07}.

From these data we extracted values for~$f_{ij}$ as described and also
calculated the full in- and out-degree sequences and used them to evaluate
the analytic expression~\eqref{eq:fij1} for the same quantity.

Figure~\ref{fig:hepthf} shows separately the value of $f_{ij}$ for fixed
$i$ and varying~$j$ (left panels) and for fixed $j$ and varying~$i$ (right
panels) for the two networks.  As we can see, in all cases the analytic
solution for the random graph model agrees surprisingly well with the
measurements.  The agreement is not perfect---there are visible differences
between measurement and theory---but the level of agreement is far better
than for most other random graph models.  Certainly the predictions of the
configuration model rarely agree this well with the behavior of real-world
networks.  Thus it appears that, in this case at least, the twin inputs of
degree sequence and vertex order are enough to capture a large part of the
variation in edge placement in the true citation networks.

There are other aspects of network structure, however, that are not so well
captured by our model.  An example is correlations between the degrees of
adjacent vertices, or degree assortativity in the nomenclature of
Section~\ref{sec:assort}.  We consider two kinds of possible degree
correlations over directed edges: correlations between in- and out-degrees
at the start and end of directed edges, and correlations between in-degrees
at either end.  In the language of paper citations, the former is a measure
of the extent to which highly cited papers are cited more often by prolific
citers.  The latter is a measure of the extent to which highly cited papers
are more likely to be cited by other highly cited papers.  We have computed
correlation coefficients of the form~\eqref{eq:corrcoefficient} for both
networks described above for both of these types of correlations, as well
as calculating expected values for random graphs with the same degree
sequences from Eq.~\eqref{eq:corrcoefficient}.

The results show mixed levels of agreement.  For the high-energy physics
citation network the measured and predicted values of the correlation
coefficients are in all cases very small, indeed negligible for most
practical purposes, so that, although the empirical and theoretical values
do not agree closely, one could claim that there is qualitative agreement
between them in that there is essentially no correlation present.  [For
in-degree/out-degree correlations we find $r=0.002$ (empirical) and
$-0.003$ (theory) and for in-degree/in-degree we find $r=0.040$ (empirical)
and $0.016$ (theory).]

For the Supreme Court, on the other hand, the correlations are more
substantial and moreover display significant disparity between observed and
predicted values.  For in-degree/out-degree correlations we find $r=0.124$
(empirical) and $0.007$ (theory), and for in-degree/in-degree we find
$r=0.184$ (empirical) and $0.022$ (theory).  This appears to indicate the
presence of significant phenomena in the real network that are not captured
in the model, and illustrates one of the main motivations for the creation
of random graph models, which is to provide a null model that can tell us
when an observed property of a network differs significantly from what we
would expect on the basis of chance, and hence draw our attention to
nontrivial network features.

\section{Conclusions}
In this paper we have introduced two random graph models for directed
acyclic graphs, which are analogous to the $G(n,m)$ and $G(n,p)$ models of
traditional random graph theory.  We have defined and calculated a number
of fundamental theoretical quantities for these models, including degree
sequences, degree distributions, edge and stub probabilities, and degree
correlations.  We have also defined the appropriate infinite-size limit of
our models and shown that a number of the central quantities of the theory
simplify in this limit.  We have compared the basic predictions of the
models with two example real-world networks, a network of citations between
physics papers and another of legal decisions, finding surprisingly good
agreement between measurement and theory for some properties, but
significant divergence in others.

Starting with the formalism developed in this paper it should be possible
to compute many other standard network quantities for random directed
acyclic graphs.  We believe that the models developed here have the
potential to shed a significant amount of light on the effects of vertex
ordering, an important defining property in many real-world networks.

\begin{acknowledgments}
  The authors thank Gavin Clarkson, Elizabeth Leicht, and an anonymous
  referee for useful input.  This work was funded in part by the National
  Science Foundation under grant DMS--0804778 and by the James S. McDonnell
  Foundation.
\end{acknowledgments}


\begin{thebibliography}{42}
\expandafter\ifx\csname natexlab\endcsname\relax\def\natexlab#1{#1}\fi
\expandafter\ifx\csname bibnamefont\endcsname\relax
  \def\bibnamefont#1{#1}\fi
\expandafter\ifx\csname bibfnamefont\endcsname\relax
  \def\bibfnamefont#1{#1}\fi
\expandafter\ifx\csname citenamefont\endcsname\relax
  \def\citenamefont#1{#1}\fi
\expandafter\ifx\csname url\endcsname\relax
  \def\url#1{\texttt{#1}}\fi
\expandafter\ifx\csname urlprefix\endcsname\relax\def\urlprefix{URL }\fi
\providecommand{\bibinfo}[2]{#2}
\providecommand{\eprint}[2][]{\url{#2}}

\bibitem[{\citenamefont{Price}(1965)}]{Price65}
\bibinfo{author}{\bibfnamefont{D.~J.~{\relax de S}.} \bibnamefont{Price}},
  \bibinfo{journal}{Science} \textbf{\bibinfo{volume}{149}},
  \bibinfo{pages}{510} (\bibinfo{year}{1965}).

\bibitem[{\citenamefont{Egghe and Rousseau}(1990)}]{ER90}
\bibinfo{author}{\bibfnamefont{L.}~\bibnamefont{Egghe}} \bibnamefont{and}
  \bibinfo{author}{\bibfnamefont{R.}~\bibnamefont{Rousseau}},
  \emph{\bibinfo{title}{Introduction to Informetrics}}
  (\bibinfo{publisher}{Elsevier}, \bibinfo{address}{Amsterdam},
  \bibinfo{year}{1990}).

\bibitem[{\citenamefont{Seglen}(1992)}]{Seglen92}
\bibinfo{author}{\bibfnamefont{P.~O.} \bibnamefont{Seglen}},
  \bibinfo{journal}{J. Amer. Soc. Inform. Sci.} \textbf{\bibinfo{volume}{43}},
  \bibinfo{pages}{628} (\bibinfo{year}{1992}).

\bibitem[{\citenamefont{Redner}(1998)}]{Redner98}
\bibinfo{author}{\bibfnamefont{S.}~\bibnamefont{Redner}},
  \bibinfo{journal}{Eur. Phys. J. B} \textbf{\bibinfo{volume}{4}},
  \bibinfo{pages}{131} (\bibinfo{year}{1998}).

\bibitem[{\citenamefont{Lehmann et~al.}(2003)\citenamefont{Lehmann, Lautrup,
  and Jackson}}]{LLJ03}
\bibinfo{author}{\bibfnamefont{S.}~\bibnamefont{Lehmann}},
  \bibinfo{author}{\bibfnamefont{B.}~\bibnamefont{Lautrup}}, \bibnamefont{and}
  \bibinfo{author}{\bibfnamefont{A.~D.} \bibnamefont{Jackson}},
  \bibinfo{journal}{Phys. Rev. E} \textbf{\bibinfo{volume}{68}},
  \bibinfo{pages}{026113} (\bibinfo{year}{2003}).

\bibitem[{\citenamefont{Jaffe and Trajtenberg}(2002)}]{JT02}
\bibinfo{author}{\bibfnamefont{A.}~\bibnamefont{Jaffe}} \bibnamefont{and}
  \bibinfo{author}{\bibfnamefont{M.}~\bibnamefont{Trajtenberg}},
  \emph{\bibinfo{title}{Patents, Citations and Innovations: A Window on the
  Knowledge Economy}} (\bibinfo{publisher}{MIT Press},
  \bibinfo{address}{Cambridge, MA}, \bibinfo{year}{2002}).

\bibitem[{\citenamefont{Fowler et~al.}(in press)\citenamefont{Fowler, Johnson,
  {Spriggs II}, Jeon, and Wahlbeck}}]{Fowler07}
\bibinfo{author}{\bibfnamefont{J.~H.} \bibnamefont{Fowler}},
  \bibinfo{author}{\bibfnamefont{T.~R.} \bibnamefont{Johnson}},
  \bibinfo{author}{\bibfnamefont{J.~F.} \bibnamefont{{Spriggs II}}},
  \bibinfo{author}{\bibfnamefont{S.}~\bibnamefont{Jeon}}, \bibnamefont{and}
  \bibinfo{author}{\bibfnamefont{P.~J.} \bibnamefont{Wahlbeck}},
  \bibinfo{journal}{Political Analysis}  (\bibinfo{year}{in press}).

\bibitem[{\citenamefont{Leicht et~al.}(2007)\citenamefont{Leicht, Clarkson,
  Shedden, and Newman}}]{LCSN07}
\bibinfo{author}{\bibfnamefont{E.~A.} \bibnamefont{Leicht}},
  \bibinfo{author}{\bibfnamefont{G.}~\bibnamefont{Clarkson}},
  \bibinfo{author}{\bibfnamefont{K.}~\bibnamefont{Shedden}}, \bibnamefont{and}
  \bibinfo{author}{\bibfnamefont{M.~E.~J.} \bibnamefont{Newman}},
  \bibinfo{journal}{Eur. Phys. J. B} \textbf{\bibinfo{volume}{59}},
  \bibinfo{pages}{75} (\bibinfo{year}{2007}).

\bibitem[{\citenamefont{Pittel and Tungol}(2001)}]{PT01}
\bibinfo{author}{\bibfnamefont{B.}~\bibnamefont{Pittel}} \bibnamefont{and}
  \bibinfo{author}{\bibfnamefont{R.}~\bibnamefont{Tungol}},
  \bibinfo{journal}{Random Structures and Algorithms}
  \textbf{\bibinfo{volume}{18}}, \bibinfo{pages}{164} (\bibinfo{year}{2001}).

\bibitem[{\citenamefont{Barak and Erd\H{o}s}(1984)}]{BE84}
\bibinfo{author}{\bibfnamefont{A.~B.} \bibnamefont{Barak}} \bibnamefont{and}
  \bibinfo{author}{\bibfnamefont{P.}~\bibnamefont{Erd\H{o}s}},
  \bibinfo{journal}{SIAM Journal on Algebraic and Discrete Methods}
  \textbf{\bibinfo{volume}{5}}, \bibinfo{pages}{508} (\bibinfo{year}{1984}).

\bibitem[{\citenamefont{McKay et~al.}(2004)\citenamefont{McKay, Oggier, Royle,
  Sloane, Wanless, and Wilf}}]{McKay04}
\bibinfo{author}{\bibfnamefont{B.~D.} \bibnamefont{McKay}},
  \bibinfo{author}{\bibfnamefont{F.~E.} \bibnamefont{Oggier}},
  \bibinfo{author}{\bibfnamefont{G.~F.} \bibnamefont{Royle}},
  \bibinfo{author}{\bibfnamefont{N.~J.~A.} \bibnamefont{Sloane}},
  \bibinfo{author}{\bibfnamefont{I.~M.} \bibnamefont{Wanless}},
  \bibnamefont{and} \bibinfo{author}{\bibfnamefont{H.~S.} \bibnamefont{Wilf}},
  \bibinfo{journal}{Journal of Integer Sequences} \textbf{\bibinfo{volume}{7}},
  \bibinfo{pages}{04.3.3} (\bibinfo{year}{2004}).

\bibitem[{\citenamefont{\L{}uczak}(1991)}]{Luczak91}
\bibinfo{author}{\bibfnamefont{T.}~\bibnamefont{\L{}uczak}},
  \bibinfo{journal}{Order} \textbf{\bibinfo{volume}{8}}, \bibinfo{pages}{291}
  (\bibinfo{year}{1991}).

\bibitem[{\citenamefont{Albert and Frieze}(1989)}]{AF89}
\bibinfo{author}{\bibfnamefont{M.~H.} \bibnamefont{Albert}} \bibnamefont{and}
  \bibinfo{author}{\bibfnamefont{A.~M.} \bibnamefont{Frieze}},
  \bibinfo{journal}{Order} \textbf{\bibinfo{volume}{6}}, \bibinfo{pages}{19}
  (\bibinfo{year}{1989}).

\bibitem[{\citenamefont{Bollob\'as and Brightwell}(1997)}]{BB97}
\bibinfo{author}{\bibfnamefont{B.}~\bibnamefont{Bollob\'as}} \bibnamefont{and}
  \bibinfo{author}{\bibfnamefont{G.}~\bibnamefont{Brightwell}},
  \bibinfo{journal}{SIAM J. Discrete Math} \textbf{\bibinfo{volume}{10}},
  \bibinfo{pages}{318} (\bibinfo{year}{1997}).

\bibitem[{\citenamefont{Jensen}(2001)}]{Jensen01}
\bibinfo{author}{\bibfnamefont{F.~V.} \bibnamefont{Jensen}},
  \emph{\bibinfo{title}{Bayesian Networks and Decision Graphs}}, Information
  Science and Statistics (\bibinfo{publisher}{Springer},
  \bibinfo{address}{Berlin}, \bibinfo{year}{2001}).

\bibitem[{\citenamefont{Ide and Cozman}(2002)}]{IC02}
\bibinfo{author}{\bibfnamefont{J.~S.} \bibnamefont{Ide}} \bibnamefont{and}
  \bibinfo{author}{\bibfnamefont{F.~G.} \bibnamefont{Cozman}}, in
  \emph{\bibinfo{booktitle}{Proceedings of the 16th Brazilian Symposium on
  Artificial Intelligence}} (\bibinfo{publisher}{Springer-Verlag},
  \bibinfo{address}{London, UK}, \bibinfo{year}{2002}), pp.
  \bibinfo{pages}{366--375}.

\bibitem[{\citenamefont{Mengshoel et~al.}(2006)\citenamefont{Mengshoel,
  Wilkins, and Roth}}]{MWR06}
\bibinfo{author}{\bibfnamefont{O.~J.} \bibnamefont{Mengshoel}},
  \bibinfo{author}{\bibfnamefont{D.~C.} \bibnamefont{Wilkins}},
  \bibnamefont{and} \bibinfo{author}{\bibfnamefont{D.}~\bibnamefont{Roth}},
  \bibinfo{journal}{Artif. Intell.} \textbf{\bibinfo{volume}{170}},
  \bibinfo{pages}{1137} (\bibinfo{year}{2006}).

\bibitem{note2}
  In fact, it is straightforward (though tedious) to show that if $f_{ij}$
  is a product of independent quantities $a_i$ and $b_j$ as here, then
  Eq.~\eqref{eq:formab} is the only possible value these quantities can
  take.  This comes as no surprise, since there are $2n-2$ each of the
  $a$'s and $b$'s and $2n-2$ constraints imposed by the degrees of the
  vertices, so one would expect the $a$'s and $b$'s to be completely
  specified by the degree sequence alone.

\bibitem[{\citenamefont{Erd\H{o}s and R\'enyi}(1959)}]{ER59}
\bibinfo{author}{\bibfnamefont{P.}~\bibnamefont{Erd\H{o}s}} \bibnamefont{and}
  \bibinfo{author}{\bibfnamefont{A.}~\bibnamefont{R\'enyi}},
  \bibinfo{journal}{Publicationes Mathematicae} \textbf{\bibinfo{volume}{6}},
  \bibinfo{pages}{290} (\bibinfo{year}{1959}).

\bibitem[{\citenamefont{Erd\H{o}s and R\'enyi}(1960)}]{ER60}
\bibinfo{author}{\bibfnamefont{P.}~\bibnamefont{Erd\H{o}s}} \bibnamefont{and}
  \bibinfo{author}{\bibfnamefont{A.}~\bibnamefont{R\'enyi}},
  \bibinfo{journal}{Publications of the Mathematical Institute of the Hungarian
  Academy of Sciences} \textbf{\bibinfo{volume}{5}}, \bibinfo{pages}{17}
  (\bibinfo{year}{1960}).

\bibitem[{\citenamefont{Bollob\'as}(1980)}]{Bollobas80}
\bibinfo{author}{\bibfnamefont{B.}~\bibnamefont{Bollob\'as}},
  \bibinfo{journal}{European Journal of Combinatorics}
  \textbf{\bibinfo{volume}{1}}, \bibinfo{pages}{311} (\bibinfo{year}{1980}).

\bibitem[{\citenamefont{\L{}uczak}(1992)}]{Luczak92}
\bibinfo{author}{\bibfnamefont{T.}~\bibnamefont{\L{}uczak}}, in
  \emph{\bibinfo{booktitle}{Proceedings of the Symposium on Random Graphs,
  Pozna\'n 1989}}, edited by \bibinfo{editor}{\bibfnamefont{A.~M.}
  \bibnamefont{Frieze}} \bibnamefont{and}
  \bibinfo{editor}{\bibfnamefont{T.}~\bibnamefont{\L{}uczak}}
  (\bibinfo{publisher}{John Wiley}, \bibinfo{address}{New York},
  \bibinfo{year}{1992}), pp. \bibinfo{pages}{165--182}.

\bibitem[{\citenamefont{Molloy and Reed}(1995)}]{MR95}
\bibinfo{author}{\bibfnamefont{M.}~\bibnamefont{Molloy}} \bibnamefont{and}
  \bibinfo{author}{\bibfnamefont{B.}~\bibnamefont{Reed}},
  \bibinfo{journal}{Random Structures and Algorithms}
  \textbf{\bibinfo{volume}{6}}, \bibinfo{pages}{161} (\bibinfo{year}{1995}).

\bibitem[{\citenamefont{Molloy and Reed}(1998)}]{MR98}
\bibinfo{author}{\bibfnamefont{M.}~\bibnamefont{Molloy}} \bibnamefont{and}
  \bibinfo{author}{\bibfnamefont{B.}~\bibnamefont{Reed}},
  \bibinfo{journal}{Combinatorics, Probability and Computing}
  \textbf{\bibinfo{volume}{7}}, \bibinfo{pages}{295} (\bibinfo{year}{1998}).

\bibitem[{\citenamefont{Newman et~al.}(2001)\citenamefont{Newman, Strogatz, and
  Watts}}]{NSW01}
\bibinfo{author}{\bibfnamefont{M.~E.~J.} \bibnamefont{Newman}},
  \bibinfo{author}{\bibfnamefont{S.~H.} \bibnamefont{Strogatz}},
  \bibnamefont{and} \bibinfo{author}{\bibfnamefont{D.~J.} \bibnamefont{Watts}},
  \bibinfo{journal}{Phys. Rev. E} \textbf{\bibinfo{volume}{64}},
  \bibinfo{pages}{026118} (\bibinfo{year}{2001}).

\bibitem[{\citenamefont{Dorogovtsev et~al.}(2001)\citenamefont{Dorogovtsev,
  Mendes, and Samukhin}}]{DMS01a}
\bibinfo{author}{\bibfnamefont{S.~N.} \bibnamefont{Dorogovtsev}},
  \bibinfo{author}{\bibfnamefont{J.~F.~F.} \bibnamefont{Mendes}},
  \bibnamefont{and} \bibinfo{author}{\bibfnamefont{A.~N.}
  \bibnamefont{Samukhin}}, \bibinfo{journal}{Phys. Rev. E}
  \textbf{\bibinfo{volume}{64}}, \bibinfo{pages}{025101}
  (\bibinfo{year}{2001}).

\bibitem[{\citenamefont{Watts and Strogatz}(1998)}]{WS98}
\bibinfo{author}{\bibfnamefont{D.~J.} \bibnamefont{Watts}} \bibnamefont{and}
  \bibinfo{author}{\bibfnamefont{S.~H.} \bibnamefont{Strogatz}},
  \bibinfo{journal}{Nature} \textbf{\bibinfo{volume}{393}},
  \bibinfo{pages}{440} (\bibinfo{year}{1998}).

\bibitem[{\citenamefont{Holland and Leinhardt}(1981)}]{HL81}
\bibinfo{author}{\bibfnamefont{P.~W.} \bibnamefont{Holland}} \bibnamefont{and}
  \bibinfo{author}{\bibfnamefont{S.}~\bibnamefont{Leinhardt}},
  \bibinfo{journal}{J. Amer. Stat. Assoc.} \textbf{\bibinfo{volume}{76}},
  \bibinfo{pages}{33} (\bibinfo{year}{1981}).

\bibitem[{\citenamefont{Strauss}(1986)}]{Strauss86}
\bibinfo{author}{\bibfnamefont{D.}~\bibnamefont{Strauss}},
  \bibinfo{journal}{SIAM Review} \textbf{\bibinfo{volume}{28}},
  \bibinfo{pages}{513} (\bibinfo{year}{1986}).

\bibitem[{\citenamefont{Karrer and Newman}(2009)}]{Karrer09a}
\bibinfo{author}{\bibfnamefont{B.}~\bibnamefont{Karrer}} \bibnamefont{and}
  \bibinfo{author}{\bibfnamefont{M.~E.~J.} \bibnamefont{Newman}},
  \bibinfo{journal}{Phys. Rev. Lett.} \textbf{\bibinfo{volume}{102}},
  \bibinfo{pages}{128701} (\bibinfo{year}{2009}).

\bibitem{note1}
  It is quite easy to demonstrate that every acyclic graph has at least one
  ordering of its vertices such that all edges point from ``later'' to
  ``earlier'' vertices, as they do in a citation network.  Thus all acyclic
  graphs are also ordered graphs.  Unfortunately, in most cases a graph
  will have more than one such ordering and the number of orderings varies
  widely with the graph.  This means that while we can create a random
  acyclic graph on $n$ vertices for which no ordering is given by first
  generating a random ordering and then using the methods described in this
  paper to generate a graph with that ordering, the resulting graph will in
  general be sampled in a nonuniform and poorly controlled way from the set
  of all directed acyclic graphs with the given vertices.  We do not at
  present know of any way to sample uniformly from the unordered ensemble.
  Luckily, it's not something we actually want to do, since such a model
  would be inappropriate as a model of real-world acyclic networks for the
  reasons given in Section~\ref{sec:ordered}.

\bibitem[{\citenamefont{Melancon et~al.}(2001)\citenamefont{Melancon, Dutour,
  and Bousquet-Melou}}]{MDB00}
\bibinfo{author}{\bibfnamefont{G.}~\bibnamefont{Melancon}},
  \bibinfo{author}{\bibfnamefont{I.}~\bibnamefont{Dutour}}, \bibnamefont{and}
  \bibinfo{author}{\bibfnamefont{M.}~\bibnamefont{Bousquet-Melou}},
  \bibinfo{journal}{Electronic Notes in Discrete Mathematics}
  \textbf{\bibinfo{volume}{10}},
  \bibinfo{pages}{202} (\bibinfo{year}{2001}).

\bibitem[{\citenamefont{Newman}(2003)}]{Newman03c}
\bibinfo{author}{\bibfnamefont{M.~E.~J.} \bibnamefont{Newman}},
  \bibinfo{journal}{Phys. Rev. E} \textbf{\bibinfo{volume}{67}},
  \bibinfo{pages}{026126} (\bibinfo{year}{2003}).

\bibitem[{\citenamefont{Pastor-Satorras
  et~al.}(2001)\citenamefont{Pastor-Satorras, V\'azquez, and
  Vespignani}}]{PVV01}
\bibinfo{author}{\bibfnamefont{R.}~\bibnamefont{Pastor-Satorras}},
  \bibinfo{author}{\bibfnamefont{A.}~\bibnamefont{V\'azquez}},
  \bibnamefont{and}
  \bibinfo{author}{\bibfnamefont{A.}~\bibnamefont{Vespignani}},
  \bibinfo{journal}{Phys. Rev. Lett.} \textbf{\bibinfo{volume}{87}},
  \bibinfo{pages}{258701} (\bibinfo{year}{2001}).

\bibitem[{\citenamefont{Newman}(2002)}]{Newman02f}
\bibinfo{author}{\bibfnamefont{M.~E.~J.} \bibnamefont{Newman}},
  \bibinfo{journal}{Phys. Rev. Lett.} \textbf{\bibinfo{volume}{89}},
  \bibinfo{pages}{208701} (\bibinfo{year}{2002}).

\bibitem[{\citenamefont{Maslov et~al.}(2004)\citenamefont{Maslov, Sneppen, and
  Zaliznyak}}]{MSZ04}
\bibinfo{author}{\bibfnamefont{S.}~\bibnamefont{Maslov}},
  \bibinfo{author}{\bibfnamefont{K.}~\bibnamefont{Sneppen}}, \bibnamefont{and}
  \bibinfo{author}{\bibfnamefont{A.}~\bibnamefont{Zaliznyak}},
  \bibinfo{journal}{Physica A} \textbf{\bibinfo{volume}{333}},
  \bibinfo{pages}{529} (\bibinfo{year}{2004}).

\bibitem[{\citenamefont{Cohen and Newman}(1985)}]{CN85}
\bibinfo{author}{\bibfnamefont{J.~E.} \bibnamefont{Cohen}} \bibnamefont{and}
  \bibinfo{author}{\bibfnamefont{C.~M.} \bibnamefont{Newman}},
  \bibinfo{journal}{Proc. R. Soc. London B} \textbf{\bibinfo{volume}{224}},
  \bibinfo{pages}{421} (\bibinfo{year}{1985}).

\bibitem[{\citenamefont{Barab\'asi and Albert}(1999)}]{BA99b}
\bibinfo{author}{\bibfnamefont{A.-L.} \bibnamefont{Barab\'asi}}
  \bibnamefont{and} \bibinfo{author}{\bibfnamefont{R.}~\bibnamefont{Albert}},
  \bibinfo{journal}{Science} \textbf{\bibinfo{volume}{286}},
  \bibinfo{pages}{509} (\bibinfo{year}{1999}).

\bibitem[{\citenamefont{Price}(1976)}]{Price76}
\bibinfo{author}{\bibfnamefont{D.~J.~{\relax de S}.} \bibnamefont{Price}},
  \bibinfo{journal}{J. Amer. Soc. Inform. Sci.} \textbf{\bibinfo{volume}{27}},
  \bibinfo{pages}{292} (\bibinfo{year}{1976}).

\bibitem[{\citenamefont{Dorogovtsev et~al.}(2000)\citenamefont{Dorogovtsev,
  Mendes, and Samukhin}}]{DMS00}
\bibinfo{author}{\bibfnamefont{S.~N.} \bibnamefont{Dorogovtsev}},
  \bibinfo{author}{\bibfnamefont{J.~F.~F.} \bibnamefont{Mendes}},
  \bibnamefont{and} \bibinfo{author}{\bibfnamefont{A.~N.}
  \bibnamefont{Samukhin}}, \bibinfo{journal}{Phys. Rev. Lett.}
  \textbf{\bibinfo{volume}{85}}, \bibinfo{pages}{4633} (\bibinfo{year}{2000}).

\bibitem[{\citenamefont{Dorogovtsev and Mendes}(2002)}]{DM02}
\bibinfo{author}{\bibfnamefont{S.~N.} \bibnamefont{Dorogovtsev}}
  \bibnamefont{and} \bibinfo{author}{\bibfnamefont{J.~F.~F.}
  \bibnamefont{Mendes}}, \bibinfo{journal}{Advances in Physics}
  \textbf{\bibinfo{volume}{51}}, \bibinfo{pages}{1079} (\bibinfo{year}{2002}).

\bibitem[{\citenamefont{Chung and Lu}(2002{\natexlab{a}})}]{CL02a}
\bibinfo{author}{\bibfnamefont{F.}~\bibnamefont{Chung}} \bibnamefont{and}
  \bibinfo{author}{\bibfnamefont{L.}~\bibnamefont{Lu}},
  \bibinfo{journal}{Annals of Combinatorics} \textbf{\bibinfo{volume}{6}},
  \bibinfo{pages}{125} (\bibinfo{year}{2002}{\natexlab{a}}).

\bibitem[{\citenamefont{Chung and Lu}(2002{\natexlab{b}})}]{CL02b}
\bibinfo{author}{\bibfnamefont{F.}~\bibnamefont{Chung}} \bibnamefont{and}
  \bibinfo{author}{\bibfnamefont{L.}~\bibnamefont{Lu}}, \bibinfo{journal}{Proc.
  Natl. Acad. Sci. USA} \textbf{\bibinfo{volume}{99}}, \bibinfo{pages}{15879}
  (\bibinfo{year}{2002}{\natexlab{b}}).

\bibitem[{\citenamefont{Bollob\'as et~al.}(2007)\citenamefont{Bollob\'as,
  Janson, and Riordan}}]{BJR07}
\bibinfo{author}{\bibfnamefont{B.}~\bibnamefont{Bollob\'as}},
  \bibinfo{author}{\bibfnamefont{S.}~\bibnamefont{Janson}}, \bibnamefont{and}
  \bibinfo{author}{\bibfnamefont{O.}~\bibnamefont{Riordan}},
  \bibinfo{journal}{Random Structures and Algorithms}
  \textbf{\bibinfo{volume}{31}}, \bibinfo{pages}{3} (\bibinfo{year}{2007}).

\bibitem{note3}
  If the windows overlap this computation is incorrect because the number
  of possible connections is not equal to the number of in-stubs multiplied
  by the number of out-stubs.  If the windows are sufficiently small,
  however, this effect is negligible.

\end{thebibliography}
\end{document}